\documentstyle[12pt]{article}
 
\begin{document}


\newcommand{\be}[1]{\begin{equation}\label{#1}}
\newcommand{\beq}{\begin{equation}}
\newcommand{\ee}{\end{equation}}
\newcommand{\beqn}[1]{\begin{eqnarray}\label{#1}}
\newcommand{\eeqn}{\end{eqnarray}}
\newcommand{\bd}{\begin{displaymath}}
\newcommand{\ed}{\end{displaymath}}
\newcommand{\mat}[4]{\left(\begin{array}{cc}{#1}&{#2}\\{#3}&{#4}\end{array}
\right)}
\newcommand{\matr}[9]{\left(\begin{array}{ccc}{#1}&{#2}&{#3}\\
{#4}&{#5}&{#6}\\{#7}&{#8}&{#9}\end{array}\right)}
\def\simlt{\mathrel{\lower2.5pt\vbox{\lineskip=0pt\baselineskip=0pt
           \hbox{$<$}\hbox{$\sim$}}}}
\def\simgt{\mathrel{\lower2.5pt\vbox{\lineskip=0pt\baselineskip=0pt
           \hbox{$>$}\hbox{$\sim$}}}}
\def\unity{{\hbox{1\kern-.8mm l}}}
\def\tanb{\tan\beta}
\def\sinb{\sin\beta}
\def\cosb{\cos\beta}
\def\sb{s_\beta}
\def\cb{\c_\beta}
\def\epr{E^\prime}
\def\al{\alpha}
\def\ga{\gamma}
\def\Ga{\Gamma}
\def\om{\omega}
\def\OM{\Omega}
\def\la{\lambda}
\def\La{\Lambda}
\newcommand{\eps}{\varepsilon}
\def\ep{\epsilon}
\newcommand{\ov}{\overline}
\renewcommand{\to}{\rightarrow}
\def\mcirc{{\stackrel{o}{m}}}
\newcommand{\bM}{\bar M} 
\newcommand{\cM}{{\cal M}} 
\newcommand{\cO}{{\cal O}} 
\newcommand{\cW}{{\cal W}} 
\newcommand{\wx}{{\rm X}} 
\newcommand{\bx}{\bar{\rm X}} 
\newcommand{\bv}{\bar{\rm V}} 
\newcommand{\wv}{{\rm V}} 
\newcommand{\tl}{\tilde{l}} 
\newcommand{\tq}{\tilde{q}} 
\newcommand{\tuc}{\tilde{u}_c} 
\newcommand{\tdc}{\tilde{d}_c} 
\newcommand{\tec}{\tilde{e}_c} 
\newcommand{\TQ}{\tilde{Q}} 
\newcommand{\TU}{\tilde{U}}
\newcommand{\TE}{\tilde{E}} 
\newcommand{\TUC}{\tilde{U}_c} 
\newcommand{\TEC}{\tilde{E}_c} 
\newcommand{\TQC}{\tilde{Q}_c} 
%
%
\makeatletter
\newcounter{alphaequation}[equation]
\def\thealphaequation{\theequation\alph{alphaequation}}
%
\def\eqnsystem#1{
\def\@eqnnum{{\rm (\thealphaequation)}}
\def\@@eqncr{\let\@tempa\relax
\ifcase\@eqcnt \def\@tempa{& & &}
\or \def\@tempa{& &}\or \def\@tempa{&}\fi\@tempa
\if@eqnsw\@eqnnum\refstepcounter{alphaequation}\fi
\global\@eqnswtrue\global\@eqcnt=0\cr}
\refstepcounter{equation}
\let\@currentlabel\theequation
\def\@tempb{#1}
\ifx\@tempb\empty\else\label{#1}\fi
\refstepcounter{alphaequation}
\let\@currentlabel\thealphaequation
\global\@eqnswtrue\global\@eqcnt=0
\tabskip\@centering\let\\=\@eqncr
$$\halign to \displaywidth\bgroup
  \@eqnsel\hskip\@centering
  $\displaystyle\tabskip\z@{##}$&\global\@eqcnt\@ne
  \hskip2\arraycolsep\hfil${##}$\hfil&
  \global\@eqcnt\tw@\hskip2\arraycolsep
  $\displaystyle\tabskip\z@{##}$\hfil
  \tabskip\@centering&\llap{##}\tabskip\z@\cr}

\def\endeqnsystem{\@@eqncr\egroup$$\global\@ignoretrue}
\makeatother


\begin{flushright}
hep-ph/9612232 ~~~~~ INFN-FE 14/96 \\ 
November 1996 \\
\end{flushright}
\vspace{10mm}

\begin{center}
{\Large \bf More Missing VEV Mechanism \\ 
in Supersymmetric SO(10) Model }
\end{center} 

\vspace{0.3cm}
\centerline{\large Zurab Berezhiani ~ and ~ Zurab Tavartkiladze }
\vspace{4mm} 
\centerline{\it INFN Sezione di Ferrara, 44100 Ferrara, Italy,}
\centerline{and} 
\centerline{\it Institute of Physics, Georgian Academy of Sciences, 
380077 Tbilisi, Georgia} 
\vspace{1.9cm}

\begin{abstract} 
The anomalous gauge $U(1)_A$ symmetry which could emerge in 
the context of the string theories can be very useful ingredient 
towards builting the complete supersymmetric $SO(10)$ theory. 
We present an example of the $SO(10)\times U(1)_A$ model 
which provides the ``all order'' solution to 
the doublet-triplet splitting problem via the missing VEV 
mechanism -- the Planck scale corrections only can induce the 
$\mu$-term naturally of order 1 TeV. 
An interesting feature of this model is 
that all relevant GUT scale VEVs are defined by the single 
dimensional parameter in the Higgs superpotential, so that 
the $SO(10)$ symmetry should break down to the MSSM practically 
at one step, without intermediate stages. 
The colour Higgsino mediated $d=5$ operators can be 
naturally suppressed.  We also extend the model 
by implementing $U(1)_A$ as a horizontal symmetry for 
explaining the fermion mass and mixing pattern, and obtain 
a predictive texture for fermion masses. This model implies 
a moderate value of $\tan\beta$ ($\sim 6-10$) and leads to 
five predictions for the low energy observables. 
It also leads to the neutrino masses and mixing pattern that could 
naturally explain both the atmospheric and solar neutrino problems. 
In addition, a remarkable interplay of the $SO(10)$ and $U(1)_A$ 
symmetries guarantees an automatic R parity conservation at any 
order in $M_P^{-1}$, and also suppresses the Planck scale induced 
B and L violating $d=5$ operators to the needed level. 
\end{abstract}

\newpage 

\section{Introduction} 

Supersymmetric Grand Unified Theories (SUSY GUT) provide the most 
plausible possibilities to understand stability of the electroweak 
scale and the unification of the gauge couplings.
It is well known \cite{Amaldi} that in
the minimal supersymmetric standard model (MSSM)
the constants $g_{3,2,1}$ of the gauge group 
$G_{321}=SU(3)\times SU(2)\times U(1)$ join at
energies $M_{G}\sim 10^{16}$ GeV, at which scale the MSSM 
can be consistently embedded in $SU(5)$ or some larger group $G$.
This suggests a paradigm that may be some SUSY GUT (and not 
directly MSSM) emerges as a field theory limit of the 
string ``Theory of Everything'' 
which then breaks down to $G_{321}$ at the scale $M_G$. 


The main problem which emerges in SUSY GUTs is a problem
of the doublet-triplet (D/T) splitting.
The MSSM Higgs doublets $H_{u,d}$ which induce the electroweak 
symmetry breaking and fermion masses should be light 
(with mass $\sim M_W$),
while their colour-triplet partners in GUT supermultiplets should 
have masses of order of $M_X$ in order to avoid too fast proton decay.  
Another puzzle is related to the 
so-called $\mu$-problem: the theory should 
provide the superpotential term $\mu H_uH_d$ with $\mu\sim M_W$.

Presently the $SO(10)$ model is a most admired candidate for the 
grand unification \cite{so10}. 
The supersymmetric $SO(10)$ GUT, with the necessary 
superfields in representations 16, 10, 45 and 54, 
could emerge from the string theories at the Kac-Moody 
level $k_{10}=2$ or larger \cite{Dienes}. 
All standard fermion states of one family: 
$q=(u,d)$, $l=(\nu,e)$, $u^c$, $d^c$, $e^c$, and 
the `right-handed' neutrino $\nu^c$ 
fit into one irreducible representation 16 of $SO(10)$. 
Two MSSM Higgs doublets $H_u$ and $H_d$ are also embedded in 
one irreducible representation 10. 
These features provide a great possibility for constructing 
the predictive ansatzes for the fermion mass matrices 
\cite{ALS,ADHRS}. 
Another virtue of the supersymmetric $SO(10)$ model is that  
it makes possible to solve the D/T problem via the missing VEV 
mechanism which was 
originally suggested by Dimopoulos and Wilczek (DW) 
\cite{MVM} and was intensively discussed in many recent papers 
\cite{BB93,BB94,HR}.

In the present paper we try to coherently approach the 
resisitent problems in $SO(10)$ by exploiting 
a very peculiar possibility suggested by the string theory 
-- an anomalous gauge $U(1)_A$ symmetry. 
Indeed, the stringy models, in addition to the GUT 
gauge group itself, can contain also several other gauge 
group factors.
The coupling constants of all gauge groups are 
determined by the dilaton VEV: 
$1/g_a^2=k_a \langle {\rm Re}(s)\rangle$, where $k_a$ are 
the Kac-Moody levels for the corresponding gauge factor $G_a$. 
One linear combination of the possible gauge $U(1)$ factors 
can be `truly' anomalous, with nonvanishing 
trace over the charges of the matter superfields, 
while the other combinations  are rendered traceless. 
Existence of an anomalous $U(1)_A$ does not imply 
an anomaly in the original string theory. 
In the field theory limit it can be understood as a result of 
truncating the string spectrum to the particle spectrum, 
and all mixed anomalies of the matter fields can be 
effectively canceled by the Green-Schwarz mechanism \cite{GS}, 
via shift of the axion field ${\rm Im}(s)$. 
This cancellation implies that the $U(1)_A^3$ anomaly coefficient 
$C_A\propto\frac13{\rm Tr}(Q^3)$ and the mixed anomaly coefficients of
$U(1)_A$ to the other factors 
$G_a$ $(C_{10}\propto{\rm Tr}(QT_aT_a)$) and to gravity 
$(C_g\propto{\rm Tr}Q)$ should be related to the corresponding 
Kac-Moody levels as 
$C_a:C_A:C_g = k_a : k_A : k_g$.  

Therefore, in the context of the string theories 
one can consider a situation when the supersymmetric 
$SO(10)$ model is accompanied by the anomalous $U(1)_A$ symmetry. 
Then the gauge constants $g_{10}$ and $g_A$ 
of $SO(10)$ and $U(1)_A$  should be unified by the condition 
$k_{10} g_{10}^2 = k_A g_A^2 = g^2_{\rm str} $
valid at the string scale $M_{\rm str}=g_{\rm str}M_P$,  
where $M_P\sim 10^{18}$ GeV is a (reduced) Planck scale, while 
the Green-Schwarz mechanism implies that the mixed 
$U(1)_A$ anomaly coefficients are related to the Kac-Moody levels 
as $C_{10}:C_A:C_g=k_{10}:k_A:k_g$. 

As it was shown in \cite{DSW}, D-term of the anomalous $U(1)_A$ 
symmetry gets a non-zero Fayet-Iliopoulos term $\xi$ \cite{FI}: 
\be{FI}
D_A = \xi + \sum Q_i |\varphi_i|^2 , ~~~~~ 
\xi= \frac{C_g}{192\pi^2} M_{\rm str}^2 
\ee 
where the sum runs over all scalar fields $\varphi_i$ present in 
the theory with $U(1)_A$ charges $Q_i$. 
Therefore, the spontaneous breaking scale of the $U(1)_A$ symmetry 
is naturally small as compared to the Planck scale but not too small:
$\sqrt{\xi}/M_P \sim 0.1 $. 

In the literature anomalous gauge symmetry $U(1)$ was applied 
as a horizontal symmetry for explaining the fermion mass 
hierarchy, utilizing the fact that the magnitude of 
$\sqrt{\xi}/M_P$ is of the order of fermion mass ratios 
in the neighbouring families \cite{IR}. 
Recently the anomalous $U(1)_A$ symmetry was applied 
in order to justify the D/T problem solution 
in the supersymmetric $SU(6)$ model 
\cite{Gia}, 
and in the missing doublet $SU(5)$ model \cite{IMDM}. 

In the present paper we show that the idea of the anomalous 
$U(1)_A$ gauge symmetry inspired by the string theory can be 
useful also for achieving a simple `all order' solution 
to the DT splitting problem via the missing VEV mechanism 
(MVM) \cite{MVM} 
in the supersymmetric $SO(10)$ theory.

 In particular, in section 2 
we reproduce the original MVM by arangement of the $U(1)_A$ 
charges of the Higgs superfileds, which solution is stable 
against the Planck scale corrections. Even more, 
the latter can induce the order 1 TeV $\mu$-term and thus 
contribute in solving the $\mu$-problem. 
We also suggest an improved model where the proton decaying 
$d=5$ operators \cite{dim5} are strongly suppressed (section 3). 
In section 4 we implement the anomalous $U(1)_A$ symmetry 
as a horizontal symmetry between the fermion generations  
and study implications of the obtained mass textures for 
the fermion mass matrices. Interestingly, this model leads to 
the exact R parity conservation due to $U(1)_A$ charge content 
of the superfields as well as to natural suppression 
of the Planck scale cutoff $d=5$ B and L violating 
operators (sect. 5). 
Finally, in sect. 6 we briefly discuss our results.   


\section{The missing VEV SO(10)$\times$U(1)$_A$ model } 

Consider a supersymmetric $SO(10)$ model 
containing the fermion superfields $f_i\sim 16$, $i=1,2,3$, and 
the Higgs superfields in the following representations:   
$H,H'\sim 10$, $S\sim 54$, $A,B,B'\sim 45$, $C,\bar{C}\sim 16,\ov{16}$ 
and two singlets $Z$ and $X$.\footnote{
Usually the Higgs 
and fermion superfields are distinguished by introducing 
the matter parity, positive for Higgses and negative for fermions. 
In sect. 5 we show that in our model we do not need to introduce 
{\em ad hoc} the matter parity and it can emerge as an 
automatic consequence of the anomalous $U(1)_A$ charges. }   
Let us assume that $X$ has a negative $U(1)_A$ charge which 
we denote as $Q_X=-2x$. The nonzero charges of the Higgs 
superfields taken as 
\be{Q}
U(1)_A: ~~~~ Q_X=-2x, ~~~ Q_{H}=-x , ~~~ Q_{H'}=x , 
~~~ Q_C=cx, ~~~ Q_{\bar{C}}=-cx  
\ee 
while $Z,S,A,B,B'$ have vanishing $U(1)_A$ charges
(the value of $c$ will be fixed later from the phenomenological 
constraints). We also invoke two additional symmetries. 
First is a discrete $R$-symmetry ${\cal R}$ under which all above 
superfields as well as the superpotential change the sign: 
\be{Rsym}
{\cal R}: ~~~ X,Z,S,A,B,B',H,H' \to -X,Z,S,A,B,B',H,H',  ~~~~ W\to -W
\ee
Second one is another global or local symmetry $U(1)'$ under which 
only $B,B'$ and $H$ have nonzero charges: 
\be{U1B}
U(1)': ~~~~ Y_B=1, ~~~~ Y_{B'}=-1, ~~~~ Y_{H}=-1 
\ee   
while all other superfields as well as the superpotential are invariant.


The most general renormalizable Higgs superpotential allowed by 
these symmetries reads as (all $SO(10)$ indices are suppressed): 
\beqn{W}
& W_{\rm Higgs}= W_1 + W_2 + W_3 + W_4  \nonumber \\ 
& W_1 = M^2 Z + Z^3 + ZS^2 + S^3  +  SA^2  \nonumber \\ 
& W_2 = ZC\bar{C} + AC\bar{C} + ZA^2   \nonumber \\ 
& W_3 = ZBB' + SBB' + ABB' \nonumber \\ 
& W_4 = BHH' + XH'^2  
\eeqn 
where the order one constants are understood in the trilinear terms, 
and the mass parameter $M$ is of the order of GUT scale 
$M_G\simeq 10^{16}$ GeV.\footnote{In fact, the scale $M$ is the 
only ad hoc scale in the theory, 
since the magnitude of the Fayet-Iliopoulos term 
$\xi$ (\ref{FI}) is essentially determined by the Planck scale 
modulo the trace over the $U(1)_A$ charges of all superfields. 
In the context of the stringy GUT it is not easy to motivate a 
presence of dimensional parameter in the superpotential, 
and it would be highly desirable to have a realistic mechanism 
which could naturally explain the presence of the linear 
term in $W_1$. 
Perhaps one could think of a situation when there is no linear  
term in the superpotential (\ref{W}) (in this case the 
discrete ${\cal R}$ symmetry extends to the continuous 
$R$-symmetry), and it effectively emerges from the coupling 
$Z\bar{Q}^\al Q_\al$ of singlet $Z$ to some "hidden" matter 
$Q$, $\bar{Q}$ from the strongly coupled sector, 
as a result of the dynamical condensation of $\bar{Q}Q$.   
} 
We also assume that charges of the fermion superfields
$f_i$ are arranged so that they have the Yukawa couplings 
solely to $H$ (see section 4). 
Note, that the mass term $\mu H^2$ as well as couplings 
$XH^2$ or $(Z+S)H^2$ are forbidden by the $U(1)_A$ symmetry.

One has to analyze the superpotential (\ref{W}) together with the 
D-terms 
\be{D} 
g_{10}^2 \left(\sum \phi_r^\dagger T_a^{(r)} \phi_r \right)^2 + 
g_A^2 \left(\sum Q_r|\phi_r|^2 - 2x|X|^2 + \xi \right)^2 ,  
\ee
where under $\phi_r$ we imply the scalar components of all 
superfields present in the theory besides $X$ with their $U(1)_A$ 
charges $Q_r$, $T_a^{(r)}$ are the $SO(10)$ generators in the 
corresponding representations. 
We assume that the trace ${\rm Tr}Q$ 
of the $U(1)_A$ charges over all matter fields is positive 
(for the concrete model see Table 1 in sect. 4).  
It is easy to see that the theory has a supersymetry conserving 
vacuum (all F- and D-terms vanish) when the scalar $X$ gets 
a non-zero VEV $\langle X \rangle = X= \sqrt{\xi/2x}$ entirely 
from the anomalous D-term (\ref{D}). 

As for the fields $Z,S,A,B,B',C,\bar{C}$, they get nonzero 
VEVs which induce the $SO(10)$ breaking to the MSSM 
$SU(3)\times SU(2)\times U(1)$. In particular,  
$C,\bar C$ have the $SU(5)$ conserving VEVs 
($\propto |+,+,+,+,+\rangle$ in terms of eigenvalues 
of the corresponding Cartan subalgebra generators), and their  
magnitudes should be equal by the vanishing of $D_{10}$: 
$\langle C\rangle= \langle \bar{C}\rangle=C$. 
The VEVs of $S$ and $A$ then break $SU(5)$ down to 
$SU(3)\times SU(2)\times U(1)$.\footnote{ 
If the last term in $W_1$ would vanish then the VEV of $A$ would be 
$SU(5)$ invariant: $r=0$ in eq. (\ref{VEV-10}). 
}
The VEVs of $B$ and $B'$ wind towards the $B-L$ direction, 
i.e. $(15,1,1)$ in terms of the $SU(4)\times SU(2)\times SU(2)'$ 
subgroup: 
\beqn{VEV-10}
& \langle S\rangle= 
S\cdot {\rm diag}(1,1,1,-3/2 ,-3/2)\otimes {\bf 1} , \nonumber \\
& \langle A\rangle= 
A\cdot {\rm diag}(1,1,1,1+r,1+r)\otimes \sigma , \nonumber  \\
& \langle B(B') \rangle =
B(B')\cdot {\rm diag}(1,1,1,0,0)\otimes \sigma ,  
\eeqn
where 
$$ {\bf 1}= \mat{1}{0}{0}{1} , ~~~ \sigma=\mat{0}{1}{-1}{0}  $$
The magnitudes of these 
VEVs all are order $M\sim M_G$, with the accuracy of the 
$\sim 1$ coupling constants in the Higgs superpotential (\ref{W}) 
(note, the VEV of singlet $Z$ in fact plays a role of 
the mass term for other superfields). 
Certainly, uncertainties in coupling constants in (\ref{W}) 
can allow up to order of magnitude hierarchy between various 
GUT scales (e.g. $C> S$, in which case $SO(10)$ 
first breaks down to $SU(5)$ and then to the MSSM). 
Once again, this feature can justify the 
one step gauge constant unification in the $SO(10)$ model. 
Note, the superpotential is arranged in such
a way that the $B-L$ direction of $B,B'$ is not affected
by the other interaction terms. 
The presence of the last term in $W_3$ guarantees that 
no unwanted light modes appear in the theory after the 
$SO(10)$ symmetry breaking which contribution could spoil 
the  gauge couplings unification. 

As far as the magnitude of the Fayet-Iliopoulos term 
$\xi$ (\ref{FI}) is essentially determined by the 
value of mixed $U(1)_A$ - gravity anomaly $C_g$,  
the scale $X=\sqrt{\xi/2x}$ in fact cannot arbitrary:  
modulo the factor $({\rm Tr}Q/2x)^{1/2}$ we have 
$X \sim 10^{17}$ GeV. In the following the latter value will 
be used for numerical estimates,  
and thus for the ratio $\eps=X/M_P$ we take $\eps\sim 1/10-1/20$.  
This estimate implies that ${\rm Tr}Q$ has a 
moderate value in units of $Q_x$ which is indeed the case 
for the model considered below. 
For the $SO(10)$ symmetry breaking VEVs we adopt the standard value 
$M_G \simeq 10^{16}$ GeV, neglecting possible split betwen their 
values and the related threshold corrections. Thus, in estimates
we take $\eps_G=M_G/M_P\sim 10^{-2}-10^{-3}$.

After substituting the relevant VEVs in $W_4$, 
mass matrices of the doublet and triplet fragments in $H,H'$ 
respectively get the form:
\be{MVM-DT}
\begin{array}{cc}
 & {\begin{array}{cc} D & \,\,\, D' \end{array}}\\ 
\vspace{2mm}
\cM_D = \begin{array}{c}
\bar{D} \\ \bar{D}' \end{array}\!\!\!\!\!&{\left(\begin{array}{cc}
0 & 0 \\ 0 & M_{22} \end{array}\right)} 
\end{array}  , ~~~~~ 
\begin{array}{cc}
 & {\begin{array}{cc} T & \,\,\, T' \end{array}}\\ 
\vspace{2mm}
\cM_T = \begin{array}{c}
\bar{T} \\ \bar{T}' \end{array}\!\!\!\!\!&{\left(\begin{array}{cc}
0 & M_{12} \\ -M_{12} & M_{22} \end{array}\right)} 
\end{array}  
\ee 
with $M_{22}\sim X$ and $M_{12}\sim B\sim M_G$. 
Therefore, all triplets in $H,H'$ are massive while the 
doublets $D$ and $\bar{D}$ contained in $H$ remain massless 
and can be identified with the MSSM Higgses $H_u$ and $H_d$ 
respectively.

Note, that our mechanism for the D/T splitting is stable 
against Planck scale corrections. The worst thing the Planck 
scale cutoff higher dimensional operators could do is 
to generate the $\mu$-term of the needed size. 

Indeed, 
the lowest order operators which could violate the DW 
pattern in the $H-H'$ mixing term are the following: 
\be{nonr}
\frac{{\rm Tr}(AB)}{M^2_{Pl}}\, (Z+S+A)HH' , ~~~~
\frac{Z+S}{M^2_{Pl}}\, (BB') (C\bar C) 
\ee 
The first operator directly induces the off-diagonal entries of order 
$\sim M_G^3/M_{Pl}^2\sim 10^{10}$ GeV in the matrix $\cM_D$. 
The second one (with combinations in brackets taken in 
the 210-channel of $SO(10)$) gives the same order contribution 
via the coupling $BHH'$ since it 
affects the DW structure of $B,B'$ VEVs. Indeed, 
after substituting the `basic' VEVs (\ref{VEV-10}) it reduces to the 
coupling containing the $(1,1,3)$ fragment of $B$ (or $B'$) linearly 
and thus the small VEV 
will be induced also on the $T_R$ direction. 
In other words, the VEV of $B$ will change to the form 
$\langle B \rangle = B\cdot {\rm diag} (1,1,1,\kappa,\kappa)$ 
with $\kappa \sim (M_G/M_{Pl})^2 \sim 10^{-6}$. 
As a result, the matrix $\cM_D$ will get off-diagonal entries 
$m\sim 10^{10}$ GeV, and thus the `seesaw' mixing of doublets 
$D,\bar{D}$ to the massive ones $D',\bar{D}'$ will generate  
the $\mu$-term $\mu H_uH_d$ just of the needed order:
$\mu\sim m^2/X \sim 1$ TeV. 

However, such MVM has a generic problem related to 
the baryon number violating $d=5$ operators \cite{dim5}. 
Indeed, if $M_{22} \sim X \sim 10^{17}$ GeV, 
then the triplets $T,\bar{T}$ are too light 
($M_T \sim M_G^2/X \sim 10^{15}$ GeV). Therefore, the cutoff scale  
of the relevant $d=5$ operators is 
\be{dim5}
(\cM_T^{-1})_{11} = 
\frac{M_{22}}{M_{12}^2} \sim\frac{X}{M_G^2}   
\ee 
and thus they mediate unacceptably fast proton decay \cite{Nath}  
which is excluded by the present experimental data \cite{PDG}, 
especially for large $\tan\beta$ which is typical for the 
$SO(10)$ models. 

Certainly, one could lower the value of $M_{22}$ to about 
$10^{14-15}$ GeV by taking very small ($\sim 10^{-2}-10^{-3}$) 
coupling constant in the last term in $W_4$ (\ref{W}).  
Alternatively, one could replace it by the higher order operator 
say $(X^3/M_P^2)H'^2$. In this case the proton lifetime 
could be acceptable. The drawback of this situation would be 
that in this case the $\mu$ term induced by the couplings 
(\ref{nonr}) would also increase by about three orders of 
magnitude, up to $10^6$ GeV, unless it is suppressed by 
very small coupling constants in the terms (\ref{nonr}). 

In the next section we propose a more appealing possibility which 
does not suffer from this problems. It naturally suppresses 
the proton decay and at the same time naturally leaves the 
$\mu$ term in the 1 TeV range.

\section{Suppressing proton decay}

We employ a proposal by Babu and Barr \cite{BB93} to use 
additional 45-plets $R$ having VEV towards the $T_R$ direction 
of $SU(4)\times SU(2)\times SU(2)'$.\footnote{
Another possibility of the stabilizing proton 
by implementing the 45-plet with VEV towards $T_R$ 
direction in the Yukawa sector was suggested in refs. \cite{Dvali}. 
}

We assume that $H$ is a 10-plet Higgs having the Yukawa couplings 
with the fermions $f_i$, while the theory includes also 
three additional 10-plet Higgses $H_{1,2,3}$. We also  
introduce two additional 45-plets $R,R'$, and prescribe the 
following charges to the states: 
\beqn{Q-new}
&& U(1)_A: ~~~~ Q_X=-2x , ~~~ 
Q_{H},Q_{H_2}=-x , ~~~ Q_{H_1},Q_{H_3}= x , 
\nonumber \\ 
&& U(1)': 
~~~~ Y_B,Y_R=1, ~~~ Y_{B'},Y_{R'}=-1, ~~~ Y_{H_1}= -1  
\eeqn
In order to distingusih Higgses $R,R'$ from $B,B'$, we also 
introduce an additional discrete symmetry  $Z'_2$ 
which changes the sign of $R,R',H_2,H_3$ while other 
superfields are invariant (for the complete charge content 
of the theory see below, Table 1).


The Higgs superpotential of the superfields $Z,S,A,B,B'C,\bar{C}$ 
still has a form (\ref{W}) but now we add also the following terms: 
\be{W3'}
W'_3 = ZRR' + SRR' + ARR'
\ee
while the last term $W_4$ in (\ref{W}) is modified as follows: 
\be{W4}
W_4 = BHH_1 + RH_1H_2 + (Z+S+A)H_2H_3 + XH_3^2 
\ee 

From the superpotential $W'_3$ one can see that there is 
a solution when the 45-plets 
$R,R'$ get the VEVs towards the $T_R$ direction: 
\footnote{ Actually for the consistency of the F-terms 
minimization one has to introduce another singlet $Z'$ with the 
same quantum numbers as $Z$ and include it in all terms of 
superpotential. 
}
\be{R}
\langle R(R')\rangle =R(R')\cdot{\rm diag}(0,0,0,1,1)\otimes\sigma 
\ee

One can take into account also the higher order Planck scale 
cutoff operators. Then the mass matrices for the doublet and 
triplet fragments in $H$ and $H_{1,2,3}$ gets the form: 
\be{M-D}
\begin{array}{cccc}
 & {\begin{array}{cccc} \,
\bar{D} ~~~~~ & ~ \bar{D}_1 ~~~~~ & ~ \bar{D}_2 ~~~~~~~ & ~~\bar{D}_3
\end{array}}\\ \vspace{3mm}
\cM_D= \begin{array}{c}
D \\ D_1 \\ D_2 \\ D_3  \end{array}\!\!\!\!\! 
&{\left(\begin{array}{cccc}
0 & O(m) & 0 & O(m) \\
O(m) & O(m') & R & O(m') \\ 
0 & -R & 0 & Z+S+A \\
O(m) & O(m') & Z+S-A & X 
\end{array}\right)} 
\end{array} 
\ee 
\be{M-T}
\begin{array}{cccc}
 & {\begin{array}{cccc} \,
\bar{T} ~~~~~~ &  ~\bar{T}_1 ~~~& ~~~~~ \bar{T}_2 ~~~~~ & ~~~~~~ \bar{T}_3
\end{array}}\\ \vspace{3mm}
\cM_T= \begin{array}{c}
T \\ T_1 \\ T_2 \\ T_3  \end{array}\!\!\!\!\! 
&{\left(\begin{array}{cccc}
0 & B & 0 & O(m) \\
-B & O(m') & O(m) & O(m') \\ 
0 & O(m)  & 0 & Z+S+A \\
O(m) & -O(m') & Z+S-A & X 
\end{array}\right)} 
\end{array} 
\ee 
Note, that all zero elements in this expression are ``all order'' 
zeros in $X/M_P$, since the $U(1)_A$ charges of these terms are 
negative. 
The possible small but non-zero entries behind the order sign 
$O$ can come form the  higher order operators 
cutoff by the Planck scale. 
The entries $O(m)$, with $m\sim 10^{10}$ GeV 
come from the operators (\ref{nonr}) considered in the previous 
section and analogous operators for $R,R'$. 
Other entries with $m'\sim \eps_G^2 X\sim 10^{11}$ GeV 
can be induced from the terms like 
\beqn{NR}
& \frac {R^2B'}{M_P^2}HH_1,~~~~
\frac {1}{M_P^2}(RB'+R'B)(Z+A+S)HH_3,~~~~\nonumber\\ 
& \frac {X}{M_P^2}(B^2+R^2)H_1^2,  
~~~~~~~\frac {Tr(AR)}{M_P^2} (Z+A+S)H_1H_2,~~~~~~
\frac {XR}{M_P^2}(Z+A)H_1H_3 \nonumber \\ 
& \frac{1}{M_P^2} (B^2R'^2+B'^2R^2+BB'RR')(Z+S+A)
\eeqn

From (\ref{M-D}) we see that the Planck scale corrections 
induce the $\mu$ term for the doublets $H_{u,d}$ 
contained in $H$, $\mu\sim m^2/X\sim 1$ TeV. 
One the other hand, now the proton becomes long living. 
If one neglects the Planck scale induced terms, then 
$(\cM_T)_{11}^{-1}$ is vanishing and thus the color 
Higgsino mediated $d=5$ operators cannot destabilize it. 
The Planck scale corrections can induce the $d=5$ operators 
cutoff by the scale: 
\be{dim5-new}
(\cM_T^{-1})_{11} \sim \frac{X}{M_P^2}  
\ee 
and thus the proton decay is extremely suppressed, 
by factor $(M_P^2/XM_G)^2\sim 10^6$ as compared to the 
standard estimates in the minimal supersymmetric $SU(5)$ model 
\cite{Nath}, and can be hardly observable even for large 
$\tan\beta\sim 100$ and low SUSY breaking masses. 


\section{Incorporating the fermion masses}

Here we attempt to incorporate the anomalous $U(1)_A$ symmetry 
also as a horizontal symmetry between the fermion generations, 
in the spirit of the earlier proposals suggested in the framework  
of the MSSM \cite{IR} or supersymmetric $SU(5)$ model \cite{IMDM}.    

Let us take as a basis the model considered in the previous section, 
and prescribe the generation dependent $U(1)_A$ charges to 
three fermionic 16-plets $f_{1,2,3}$ so that only the third 
family $f_3$ is allowed to have the renormalizable 
Yukawa coupling $f_3 f_3 H$ to the Higgs $H$, while  
the other fermions can get masses from the higher order operators 
including powers of $X/M_P$. These operators in the superpotential 
can be effectively induced after integrating out some heavy 
states with masses $\sim M_P$ \cite{FN}.   
Namely, let us take $Q(f_i)= \frac12 x + 2(3-i)x$, and also 
assume that under the discrete ${\cal R}$ symmetry (\ref{Rsym}) 
the fermions transform as $f_i \to (-1)^{i} f_i$  
while with respect to remaining symmetry groups they are invariant. 
We also introduce an additional fermion state $F\sim 10$ with 
$Q_F=2nx$ (integer $n$) and fix the $U(1)_A$ charges 
of $C,\bar{C}$ in (\ref{Q}) as $c=-(4n+1)/2$. 
The superfield representation and charge content for the case 
$n=1$ (hereafter to be referred as to Model 1) is given in Table 1. 
\footnote{
In principle fermions could have non-trivial charges with respect 
to $U(1)'$ symmetry, as well as they could transform nontrivially 
with respect to $Z'_2$ symmetry which also could be extended 
to $U(1)''$ symmetry related to independent rotations of the 
$R,R'$ states. Then the interplay of these symmetries could fix some 
interesting textures for the quark and lepton mass matrices, 
in the spirit of the works \cite{ADHRS}. However, in the 
present paper we take more modest approach extending 
for fermions only the anomalous $U(1)_A$ symmetry. 
}

\begin{table}
\caption{The superfield transformation properties in the 
Model 1 ($n=1$). With respect to ${\cal R}$ symmetry all superfields 
change the sign except $f_2$ which is invariant. } 
\label{tab1}
$$\begin{array}{|c|cccccccccccccccc|}
\hline \vspace{1mm}
& X & Z & S & A & B & B' & R & R' & C & \ov{C} 
& H_3 & H_2 & H_1 & H & f_i & F \\
\hline SO(10)&1 &1 &54 &45 &45 &45 &45 & 45 & 16 & \ov{16} 
&10 &10 & 10 &10 &16 &10 \\
\hline U(1)_A & -2 &0 & 0 & 0 & 0 & 0 & 0 & 0 &-\frac{5}{2}
&\frac{5}{2} & 1 & -1 & 1 & -1 & \frac{13}{2}-2i & 2 \\ 
\hline
U(1)'& 0 & 0 & 0 & 0 & 1 & -1 & 1 & -1 & 0 & 0 & 0 & 0 & -1 & 0 & 0 & 0 \\ 
\hline
Z'_2 & + & + & + & + & + & + & - & - & + & + & - & - & + & + & + & +  \\ 
\hline
\end{array}$$
\end{table}


Then the possible Yukawa terms in the superpotential 
can be expressed as: 
\beqn{Yuk} 
& W_{\rm Yuk} = \cW_1 + \cW_2 + \cW_3 ,  \nonumber \\ 
& \cW_1 = g_{ij} Hf_if_j  \left(\frac{X}{M_P}\right)^{6-i-j} , 
\nonumber \\  
& \cW_2 = \gamma_{ij} \frac{\bar{C}\bar{C}}{M_P} f_if_j 
\left(\frac{X}{M_P}\right)^{2n+7-i-j}   \\  
& \cW_3 =  h_{i} Cf_i F \left(\frac{X}{M_P}\right)^{3-i} 
+ (hZ+h'S) F^2 \left(\frac{X}{M_P}\right)^{2n} 
\nonumber 
\eeqn 
where $g_{ij}=g_{ji}$, $\ga_{ij}=\ga_{ji}$ and $h_i$ ($i=1,2,3$) 
are order 1 constants. 

   There is another relevant coupling which has to be taken into 
consideration. The symmetries of the theory allow the 
following term in the Higgs superpotential: 
\be{HCC}
H \bar{C}\bar{C} \left(\frac{X}{M_P}\right)^{2n}  
\ee
while no term $HCC$ is allowed since it has a negative 
$U(1)_A$ charge. 

Let us use the $SU(5)$ subgroup language to describe their 
contributions. The decomposition of the relevant 
representations is $16=\bar5+10+1$, $\ov{16}=5+\ov{10}+1$ and  
$10=\bar5+5$. 

The coupling (\ref{HCC}) induces the mixing 
between the Higgs states $\bar5_H$ and $\bar5_C$ 
in $H$ and $C$. Therefore, the physical light Higgs $H_d$ 
does not come entirely from $H$, but is living partially 
also in $\bar5_C$, with a weight $w \sim \eps^{2n}$. 
As for the another MSSM Higgs $H_u$, it comes entirely from 
$5_H$ without any admixture from $5_{\bar{C}}$.

It is clear how each of the terms (\ref{Yuk}) is functioning. 
The first term $\cW_1$ induces the $SO(10)$ invariant mass entries 
for quarks and leptons. In the $SU(5)$ language it 
reduces to the couplings 
\be{W1}
 \cW_1 \to 
\la_{ij} 10_i 10_j 5_H + \la_{ij} 10_i \bar5_j \bar5_H + 
\la_{ij} \bar5_i 1_j 5_H ,  
~~~~~~~ \la_{ij}=g_{ij}\eps^{3-i-j}  
\ee

The second term in (\ref{Yuk}), after substituting 
the $SU(5)$ invariant VEV $C$ generates the Majorana mass 
matrix $M_{ij}$ for the $SU(5)$ singlet `right-handed' neutrinos 
$\nu^c_i=1_i$ in $f_i$: 
\be{W2}
\cW_2 \to M_{ij} \nu^c_i \nu^c_j,  
~~~~~~~~~~~~~~~~ M_{ij} = \ga_{ij}M_P\eps_C^2\eps^{2n+7-i-j}, 
\ee

The first term in $\cW_3$ induces the $SU(5)$ invariant 
couplings between the $5_F$ and $\bar5_F$ states in $F$ with 
the relevant fragments in $f_i$, which allows to avoide the 
proportionality of the down quarks and charge leptons mass matrices 
to the mass matrix of upper quarks $\sim \la_{ij}$. 
The last coupling in $\cW_3$ induces masses for the 
$5_F$ and $\bar5_F$ states in $F$. As far as $S$ contains the 
24-plet of $SU(5)$ which VEV breaks the quark-lepton symmetry, 
the mass terms for the doublet ($l$-type) and triplet ($d^c$-type) 
are different: 
\beqn{W3}
&& \cW_3 \to h_i 10_i \bar5_F \bar5_C\eps^{3-i} 
+ h_i 5_F\bar5_i C\eps^{3-i} + \eps^{2n} M_{1+24} 5_F\bar5_F , \\
&& ~~~~~~~~~~~~~~~~~~
M_{1+24}={\rm diag}(Z+S,Z+S,Z+S,Z-\frac32 S, Z-\frac32 S)
\nonumber 
\eeqn 

The VEVs of the MSSM Higgs doublets 
$\langle H_u \rangle=v_u$ and $\langle H_d \rangle=v_d$ 
($v_u^2+v_d^2=v^2$, $v=174$ GeV, and $v_u/v_d=\tanb$) 
break the electroweak symmetry and induce the fermion masses. 
In the following we have to take into account that the 
`down' VEV $v_d$ dominantly comes from $\bar{5}_H$ 
while the state $\bar{5}_C$ also has a VEV $v'_d=w v_d$, 
$w \sim \eps^{2n}$. 
At the MSSM level the first term in (\ref{W1}) reduces 
to the Yukawa couplings $\la_{ij}q_iu^cH_u$. 
The upper quark masses emerge solely from this term 
and thus $m^u_{ij}=\la_{ij} v_u$ is the mass matrix of the 
upper quarks. 
Without lose of generality, one can rotate all 
16-plets $f_i$ so that the matrix $\la_{ij}$ is diagonal: 
$\la_{ij}= {\rm diag}(\la_u,\la_c,\la_t)$. These rotations 
with angles $\sim \eps$ lead only to irrelevant 
redefinition of the other parameters without changing their 
hierarchial pattern. 

So, we take the basis where the upper quark mass 
matrix is diagonal: 
\be{m-u}
\begin{array}{ccc}
 & {\begin{array}{ccc} \!\!\!\!\!\! u^c_1 ~ &  u^c_2 ~  &  u^c_3 ~
\end{array}}\\ \vspace{2mm}
\hat{m}^u= \begin{array}{c}
u_1 \\ u_2 \\ u_3  \end{array}\!\!\!\!\! &{\left(\begin{array}{ccc}
\la_u & 0 & 0 \\ 
0 & \la_c & 0 \\ 
0 & 0 & \la_t \end{array}\right)} 
\cdot v_u 
\end{array} 
\ee 
where the Yukawa eigenvalues scale as 
$\la_t\sim 1$, $\la_c\sim \eps^2$ and $\la_u\sim \eps^4$, 
which for $\eps\sim 1/10-1/20$ properly fit the up 
quark mass pattern.

As far as $X$ is a $SO(10)$ singlet, 
the operators $\cW_1$ lead to exactly the same contribution 
to the charged lepton and the down quark mass Yukawa terms 
as well as to the neutrino Dirac terms: 
$\la_{ij}(q_id^c_jH_d + e^c_il_jH_d + \nu^c_il_j H_u)$. 
Hence, in the absence of the other contributions we would have 
$\hat{m}^e=\hat{m}^d=\hat{m}^u/\tanb$,  
and thus $m_{e,\mu,\tau}=m_{d,s,b}$ in the $SU(5)$ limit. 
It is clear that the additional fermion 10-plet $F$ 
was introduced in order to remove the degeneracy between 
the upper quark, down quark and charged lepton states. 
In the $SU(5)$ fragments $F=5_F+\bar5_F$ are contained 
only the $d^c$ and $l$ type particles and their 
conjugates. Moreover, since the 54-plet $S$ contains 24-plet 
of $SU(5)$, its VEV removes the mass degeneracy between down quark 
$d^c$ and lepton $l$ states.  
Therefore, mixing of the corresponding fermion states in 16-plets 
$f_i$ 
would induce different Clebsches for the down quark and lepton 
mass entries and thus can remove the GUT scale $SU(5)$ 
degeneracy of the physical states $s-\mu$ and $d-e$.


Hereafter we concentrate on the case $n=1$ (Model 1), leaving 
another possibilities for the future study (the models with 
$n=0$ or 2 also could be of phenomenological interest, 
but the cases with larger $n$ seem not very appealing).  
Then the big mass entries which emerge from $\cW_3$ read as 
$d_F(M_3 d^c_F + h_1\eps^2Cd^c_1 + h_2\eps Cd^c_2 + h_3C d^c_3)$ 
and 
$l^c_F(M_2 l_F + h_1\eps^2Cl_1 + h_2\eps Cl_2 + h_3Cl_3)$, 
where $M_{2,3}\sim \eps^2 M_G$ are respectively the doublet 
and triplet mass entries in $F$ induced by the last term 
in (\ref{W3}). Therefore, after decoupling the superheavy 
$5_F+\bar5'$ state where $\bar5'$ is dominantly contributed by 
$\bar5_3$, we see that the light states are left in 
$\bar5'_1\simeq \bar5_1$,  $\bar5'_2\simeq \bar5_2$ 
and $\bar5'_F\simeq \bar5_F$ which contain $\bar5_3$ as 
small admixtures. Notice also that the doublet and triplet 
fragments in $\bar5_F$ are contained in $\bar5'$ with 
different weights, $s_e=M_2/h_3C$ and $s_d=M_3/h_3C$, 
which will be used for removing the $SU(5)$ mass degeneracy 
between the down quarks and charged leptons.

Since the superheavy state is formed essentially by 
$5_F$ and $\bar5_3$, the mass matrix of the down 
quarks and charged leptons should be strongly altered by 
the rearrangement of the $\bar5$ states.  Taking also into account that 
the MSSM Higgs doublet partially (with the weight $w\sim \eps^2$) 
resides in $\bar5_C$, 
after decoupling the heavy fermions \cite{FN} we obtain the 
following mass matrices for the down quarks and charged leptons: 
\be{m-de}
\begin{array}{ccc}
 & {\begin{array}{ccc} \,d'^c_1 & \,\,~~ d'^c_F & \,\,~~ d'^c_2
\end{array}}\\ \vspace{3mm}
\hat{m}^d= \begin{array}{c}
d_1 \\ d_2 \\ d_3  \end{array}\!\!\!\!\! &{\left(\begin{array}{ccc}
\la_u{\rm e}^{{\rm i}\omega} & \eps_1\eps_2\la & 0 \\ 
\eps_1\la_c  & \eps_2\la & \la_c  \\ 
0 & \kappa_d & \eps_2\la_t \end{array}\right)} 
\end{array}  \!\! \cdot v_d, ~~~
\begin{array}{ccc}
 & {\begin{array}{ccc} \,l'_1 ~& \,\,~~ l'_F ~~& \,\,~~ l'_2
\end{array}}\\ \vspace{3mm}
\hat{m}^e= \begin{array}{c}
e^c_1 \\ e^c_2 \\ e^c_3 \end{array}\!\!\!\!\! &{\left(\begin{array}{ccc}
\la_u{\rm e}^{{\rm i}\omega} & \eps_1\eps_2\la & 0 \\ 
\eps_1\la_c  & \eps_2\la & \la_c  \\ 
0 & \kappa_e & \eps_2\la_t \end{array}\right)} 
\end{array}  \!\! \cdot v_d, 
\ee 
and the matrix of the neutrino Dirac masses: 
\be{m-nu}
\begin{array}{ccc}
 & {\begin{array}{ccc} \,\nu'_1 \, & \,\,~~ \nu'_F & \,\,~~ \nu'_2
\end{array}}\\ \vspace{2mm}
\hat{m}_D= \begin{array}{c}
\nu^c_1 \\ \nu^c_2 \\ \nu^c_3  \end{array}\!\!\!\!\! 
&{\left(\begin{array}{ccc}
\la_u{\rm e}^{{\rm i}\omega} & 0 & 0 \\ 
\eps_1\la_c  & 0 & \la_c  \\ 
0 & \kappa_e\! -\!\la  & \eps_2\la_t \end{array}\right)} 
\end{array}  \!\! \cdot v_u, 
\ee 
where the parameters are the following: 
$\eps_1=\eps h_1/h_2\sim \eps$, $\eps_2=\eps h_2/h_3\sim \eps$ 
and $\la=h_3w \sim \eps^2$. We see that all entries 
are the same in the matrices $\hat{m}^{d,e}$  except the 2,3 elements 
$\kappa_d=\la-\la_t s_d$ and $\kappa_e=\la-\la_t s_e$ 
which are generally different. The impact of the latter 
is to remove the troblessome $SU(5)$ degeneracy between 
the $s-\mu$ and $e-d$ mass eigenvalues at the GUT scale.
Without lose of generality, all entries in $\hat{m}^{d,e}$ 
can be chosen real by redefinition of the fermion phases 
except the 1,1 entry and 2,3 entries $\kappa_{d,e}$ which 
remain complex. 

Mass matrices (\ref{m-u}), (\ref{m-de}) and (\ref{m-nu})
depend on 12 parameters, consisting of the Yukawa constants 
$\la_{u,c,t}$, $\tanb=v_u/v_d$  
and 8 unknown parameters: $\eps_{1}$, $\eps_2$, $\la$, 
the phase $\omega$ and two complex parameters $\kappa_{e,d}$. 
Therefore, in general we have to obtain 2 relations between the 14 
observables of the MSSM (nine fermion masses,  
four parameters of the CKM matrix: $s_{12}=|V_{us}|$, 
$s_{23}=|V_{cb}|$, $s_{13}=|V_{ub}|$ and $CP$-phase $\delta$,  
and still $\tanb$). 
The general analysis of these mass matrices will be presented 
elsewhere. For the moment we confine ourselves by illustrating 
the particular case when $\kappa_{e,d}$ are real 
and $\kappa_d=-\kappa_e=\kappa$. In these case the number of 
parameters are reduced by three and hence one has to obtain 
five predictions. 

Indeed, neglecting the $O(\eps)$ corrections in diagonalization 
of the matrices $\hat{m}^{d,e}$, we obtain 
the following relations holding with some 10 percent accuracy: 
\beqn{results}
&& \la_b=\la_\tau, ~~~~~ \la_c=s_{23}\la_\tau=\frac23\la_\mu, 
~~~~~ \la_s=\frac13 \la_\mu,  
\la_e=|\frac13 \la_d + \la_u{\rm e}^{i\omega}|, ~~~~~
\nonumber \\
&& 
\sin\delta=\frac{\la_u}{\la_d}s_{12}\sin\omega, ~~~~~~
s_{13}=\frac{\sqrt{\la_d\la_s}}{2\la_b}=\frac14 s_{12}s_{23} , 
\eeqn 
while for the values of the unknown parameters we have: 
\beqn{par}
&& \eps_2\la=\frac12 (\la_\mu+\la_s)=\la_c , ~~~~~
\kappa=\frac{\la_\mu-\la_s}{2\la_c}\la_\tau = \frac12\la_\tau 
\nonumber \\ 
&& \eps_1\la_c=\eps_1\eps_2\la=\sqrt{\la_d\la_s}\simeq  
\sqrt{\la_e\la_\mu}, ~~~~ 
\eps_2=\frac{\la_t}{\la_\tau}\sim 0.1 , ~~~~
\la=s_{23}\la_t \sim \la_\tau 
\eeqn 
The first prediction $\la_b=\la_\tau$ in (\ref{results}), 
the famous $b-\tau$ Yukawa unification at the GUT scale, 
immediately follows 
from the assumption that the (3,3) element $\eps_2\la_t$ is the 
largest entry in the matrices (\ref{m-de}). 
Parametrically it is indeed $\sim \eps$ while the other entries 
should be smaller: $\la_c,\kappa_{e,d}\sim \eps^2$, etc. 
The second relation follows from the fact that 
in our model the $V_{cb}$ element of the CKM matrix 
is given as $s_{23}=\la_c/\la_\tau$. 
Then taking as input the experimental values $s_{23}=0.04$ and 
$\la_\mu/\la_\tau=0.06$, we obtain $\la_c=\frac23\la_\mu$.  

The three next predictions in (\ref{results}) can be derived 
by performing the diagonalization of the $1-2$ blocks in the 
matrices (\ref{m-de}) and putting the Cabibbo angle to its 
experimental value $s_{12}=0.22$. And finally, the last 
prediction for $s_{13}$ element in the CKM matrix emerges 
due to the big mixing between the right-handed 
states $d^c_F$ and $d^c_2$ in $\hat{m}^d$: 
$s_{23}^c=\kappa/\la_b=\frac12$. 
\footnote{
We see that $\kappa=1/2\eps_2\la_t\sim \eps$, in some 
contradiction with the parametrical estimate of its value 
$\kappa\sim \eps^2$. Such an enhancement should not be 
surprising and can have an accidental origin due to some 
conspiracies in the parameter space of the model.   
However, this value of $\kappa$ is still enough small to 
treat the rotation angle $s^c_{23}$ between the 
right states $d^c_F$ and $d^c_2$ as small angle, 
and it and does not affect obtained results more 
than about 10\%. One can see, that for arbitrary complex 
$\kappa_{e,d}$ their modulus always be more than above value 
which in general could spoil our approximation and, in particular, 
the $b-\tau$ Yukawa unification. Therefore, our choice 
$\kappa_e$ and $\kappa_d$ as both real and opposite 
maximally corresponds to the spirit of our approximation. 
} 

One can translate the relations (\ref{results}) into predictions 
for the low energy physical observables. 
As it is well known, the $b-\tau$ Yukawa unification at the 
GUT scale explains the value of the $b$-quark mass with 
implication that top mass should be close to its infrared fixed 
value $M_t\simeq \sinb 200$ GeV. 
The second relation in (\ref{results}) can be used for deducing 
the value of $\tanb$ using the experimental value of $c$ quark 
mass. Within the uncertainties related to the experimental 
values of $\al_3(M_Z)$ and $M_t$, we obtain $\tanb\simeq 6-10$.  
The next relation fixes the $s$ quark mass: 
$m_s\simeq 130-180$ MeV, in agreement with the `current-algebra'
predictions. The fourth relation implies that $m_d/m_s\simeq 1/22$,  
with about 20 percent uncertainty $\sim(3m_u/m_d\tanb)$ 
related to $\omega$ varying from 0 and $\pi$. 
The $CP$-violating phase is 
very small -- even for $\omega=\pi/2$, it cannot exceed 
the value $\delta\sim s_{12}\la_u/\la_d\sim 0.01$. 
\footnote{In principle this is no problem 
since in the context of supersymmetric grand unified 
theory the $CP$-violation $K^0-\bar{K}^0$ system, even too strong, 
can be originated from the supersymmetric contributions to 
both the $\epsilon_K$ and $\epsilon'_K$ parameters. 
}
And finally, the last relation in (\ref{results}) implies 
that $V_{ub}/V_{cb}\simeq 0.06$, an agreement with the current 
experimental range \cite{PDG}. 


Let us now turn to the neutrino masses. 
The mass matrix of the physical left handed states 
$\nu_i$ which results from the `seesaw' decoupling \cite{seesaw} 
of the heavy Majorana states $\nu^c_i$ has a form  
$\hat{m}_\nu = \hat{m}_D^T\hat{M}^{-1} \hat{m}_D$,  
where $\hat{m}_D$ is the Dirac mass matrix of eq. (\ref{m-nu}), 
and 
$\hat{M}_{ij} = \gamma_{ij}\eps^{6-i-j} M_R$ is the Majorana 
mass matrix of the $\nu^c$ states,  
where $M_R\sim \eps_C^2 \eps^3 M_P \sim 10^{11-12}$ GeV 
corresponds to the magnitude of its heaviest eigenstate.   
The matrix $\hat{M}$ cannot be exactly fixed from the theory, 
but it should have typical structure with the 
eigenvalues having a hierarchy $\eps^4:\eps^2:1$ and with 
the rotation angles between the neighbouring families $\sim \eps$. 

Putting all these together, one can see that the following picture 
emerges for the neutrino masses and mixing. 
The neutrino mass eigenvalues exhibite approximatelly the 
hierarchy 
$m_{\nu_\tau}: m_{\nu_\mu}:  m_{\nu_e} \sim 1:\eps:\eps^2 $, 
where the mass of the heaviest ($\nu_\tau$) state 
$m_{\nu_\tau}\sim \la_\tau^2v_u^2/M_R$ 
can naturally emerge in the range of $0.1$ eV. Then 
the mass of $\nu_\mu$ can be of about $3\times 10^{-3}$ eV. 

On the other hand, there should be the strong mixing 
between the $\nu_\mu$ and $\nu_\tau$ states. 
This can be seen by comparing the Dirac mass matrix 
$\hat{m}_D$ with the parameter values calculated in (\ref{par}) 
to the charged lepton mass matrix $\hat{m}^e$ in (\ref{m-de}). 
We see that 2,3 rotation angles needed for the diagonalization 
of these matrices differ by a quantity $\sim \la/\la_\tau\sim 1$.
One can hardly imagine that this large angle will be 
cancelled by contributions from the unknown parameters 
in $\hat{M}$. Therefore, we expect that 
$\sin^2\theta_{\mu\tau}\sim 1$. 
The mixing between the $\nu_e$ and $\nu_\mu$ states is smaller. 
By comparing the matrices $\hat{m}^e$ and $\hat{m}_D$, one 
finds a contribution $\sim \sqrt{\la_e/\la_\mu}\sim \eps$, 
while the $\sim \eps$ contributions can come also from 
the structure of $\hat{M}$. Therefore, we expect that 
$\sin^2\theta_{e\mu}\sim \eps^2\sim 10^{-2}$. 

These features of the neutrino mass spectrum and mixing 
provide an appealing possibility to explain simultaneoulsy 
the atmospheric and solar neutrino problems:
deficite of the atmospheric muon neutrinos \cite{ANP} can be 
due to the $\nu_\mu - \nu_\tau$ oscillation 
with $\delta m^2\sim 10^{-2}$ eV$^2$ and $\sin^2 2\theta \sim 1$, 
while the solar neutrino problem can be explained by the 
MSW oscillation $\nu_e - \nu_\mu$ \cite{MSW}
with $\delta m^2 \sim 10^{-5}$ 
eV$^2$ and $\sin^2 2\theta \sim \eps^2 \sim 10^{-2}$.


\section{Automatic R parity }

As far as our model includes Higgses in representations 
$C,\bar{C}\sim 16,\ov{16}$ and the fermions $F\sim 10$, 
at the first glance the R parity conservation is not 
automatic anymore, 
%
%
and the low energy theory (MSSM) 
should include the B and L violating $d=3$ and $d=4$ operators: 
\be{R-SM}
\mu'_i l_i H_u + \la_{ijk} l_i l_j e^c_k + 
\la'_{ijk}l_i q_j d^c_k + \la''_{ijk}u^c_i d^c_j d^c_k    
\ee 
These could emerge from renormalizable couplings 
like $FH$, $f_iCH$, $Ff_if_j$ (recall that 10-plet $F$ is strongly 
mixed to the physical light fermion states), or nonrenormalizable 
operators like $\frac{1}{M_{Pl}}f_if_jf_kC$ or 
$\frac{1}{M_{Pl}}F^2f\bar{C}$ after substituting the VEV $C$
unless they are forbidden by an {\em ad hoc} matter parity.

Nevertheless, in our theory the R parity conservation occurs 
to be automatic due to the $U(1)_A$ symmetry: 
the $SO(10)\times U(1)_A$ invariant terms containing the odd 
number of the fermion superfields cannot emerge at any order 
in $M_{Pl}^{-1}$. 
The simple proof of this statement can be red out the Table 2. 
Indeed, any $SO(10)$ invariant operator containing the fermion 
superfields in the odd number can be presented as 
 a product $\psi\cdot \Omega_\psi$, 
where $\psi=F$, $f_iC$ or $f_i\bar{C}$, and $\Omega_\psi$ is a 
complementary operator in the same $SO(10)$ representation 
as $\psi$, built upon the Higgses and even number of fermions. 
More precisely, 
$\Omega_\psi= \Omega_{\rm ferm}\cdot\Omega_{\rm Higgs}$, 
where a tensor $\Omega_{\rm ferm}$ combines an even number 
of fermions and a tensor $\Omega_{\rm Higgs}$ consists 
entirely of Higgses. 

One can characterize all these tensors by the following 
two parities: 

$\bullet$ the $SO(10)$ parity $D$: negative (positive) 
for tensors with the odd (even) number of the fundamental (10-plet) 
indices; 

$\bullet$ the $U(1)_A$ parity $N$: negative (positive) 
for combinations with odd (even) value of the $U(1)_A$ charge in 
units of $x$.

From Table 1 we see that for $\psi=F,f_iC,f_i\bar{C}$ 
the $N$ parity is always opposite to the $D$-parity. 
On the contrary, for any tensor $\Omega_{\rm Higgs}$ 
the $D$ and $N$ parities always coincide: 
Higgses with positive $D$ (1,45,54 representations, among those 
combinations $C\bar{C}\sim 1,45,210$) all have positive $N$, 
and Higgses with negative $D$ (10-plets and combinations 
$CC\sim 10,120,126$) have also negative $N$. Clearly, 
the same is true for any tensor $\Omega_{\rm ferm}$ composed 
upon the even number of fermions, e.g. $\sim F^{2n}$ 
or $(f\cdot f)^n$.


\begin{table}
\caption{The $D$ and $N$ parities of various $SO(10)$ tensors. }
\label{t:tab2}
$$\begin{array}{|c|c|c|c|c|c|c|}
\hline 
& F & fC & f\overline C & F^{2n} & (f\cdot f)^n &\Omega_{\rm Higgs} \\
\hline SO(10):~ D & - & - & + & + & +~/~- & +~/~- \\ 
\hline
U(1)_A:~ N & + & + & - & + & +~/~- & +~/~-  \\
\hline
\end{array}$$
\end{table}


So, $N$ and $D$ parities of $\Omega_\psi$ always coincide, 
and for $\psi$ itself these parities are always opposite. Thus, 
the structure of the matter parity breaking operators can never 
match both the $SO(10)$ and $U(1)_A$ symmetries: 
the $SO(10)$ invariant terms containing an odd number of fermions 
are forbidden by the $U(1)_A$ symmetry,  
and only the operators with the even number of fermions 
(like the Yukawa terms in (\ref{Yuk})) can be allowed. 
In other words, the theory has an accidental matter parity $Z_2$ 
under which the fermion superfields  $f_i$ and $F$ change sign 
while the Higgs superfields are invariant.  

Hence, our $SO(10)$ model provides an attractive possibility to 
understand the Baryon and Lepton number conservation in $d=3$ and 
$d=4$ operators, without imposing matter parity (R-parity) 
in an {\it ad hoc} manner. 
The exact R-parity conservation emerges as an automatic (accidental) 
consequence of the $U(1)_A$ charge content of the fields in the theory. 
Note, the additional abelian $U(1)'$ or discrete ${\cal R}$ 
and $Z'_2$ symmetries  play no role in deriving this property. 

Let us conclude this section with the following remark. 
The $d=5$ B and L violating operators contain the even number 
of fermions and thus they can emerge in the Planck scale 
cutoff terms. 
However, in our model they are naturally suppressed by the 
by the $U(1)_A$ symmetry. 
Indeed, the family dependent fermion charges $Q(f_i)$ allow  
the relevant terms only at the following order:  
\be{dim5-MP} 
\frac{1}{M_P} f_i f_j f_k f_n  
\left(\frac{X}{M_{Pl}}\right)^{13-i-j-k-n} 
\Longrightarrow 
~ \frac{\eps^{13-i-j-k-n}}{M_P} (q_i q_j q_k l_n + 
u^c_i u^c_j d^c_k e^c_n)
\ee 
Consider e.g. the dangerous term $q_1q_1q_2l_2$ 
leading to the decay $p\to K^+ \nu_\tau$ 
(recall that the state $l_3\subset f_3$ in our model is superheavy 
and the third generation of leptons actually comes from $f_2$). 
We see that its constant is suppressed by factor $\sim \eps^7$ 
which for $\eps\sim 1/10-1/20$ 
can be enough to rise the proton lifetime above 
the experimental limits.

\section{Discussion } 

We find that the anomalous gauge $U(1)_A$ symmetry can be of 
great help for builting the complete supersymmetric $SO(10)$ model. 
It could emerge together with the $SO(10)$ gauge group 
in the string theory context, and play a key role in 
in solving various SUSY GUT puzzles as are 
the gauge hierarchy and doublet-triplet splitting problem, 
problem of fermion mass hierarchy, origin of matter parity 
(or R parity) conservation and so long lifetime of proton.  
In particular, we have shown some examples of supersymmetric 
$SO(10)\times U(1)_A$ models which could provide an 
``all order'' stable solution to the D/T problem via the 
missing VEV mechanism. 
We have also extended a picture for the fermion masses 
by involving $U(1)_A$ as a horizontal symmetry. 
The fermion mass hierarchy as well as the magnitudes 
of the CKM mixing angles can be naturally understood in 
terms of small parameter ($\eps\sim 1/10-1/20$) 
with a proper choice of the fermion $U(1)_A$ charges. 
In addition, 
the $U(1)_A$ charge content of superfields 
in the theory can be arranged so that R parity breaking 
operators will be forbidden at any order in $M_P^{-1}$. 
In other words, the exact conservation of R parity  
can be an accidental consequence of the gauge symmetry. 
The suggested pattern for the neutrino masses and 
mixing can be of phenomenological interest. 



\begin{thebibliography}{9}

\bibitem{Amaldi}  
U. Amaldi, W. de Boer and H. Furstenau, Phys. Lett. B 260 (1991) 447; \\
J. Ellis, S. Kelley and D. Nanopoulos, Phys. Lett. B 260 (1991) 131; \\
P. Langacker and M. Luo, Phys. Rev. D 44 (1991) 817. 


\bibitem{so10} 
H. Fritzsch and P. Minkowski, Ann. Phys. 93 (1975) 193; \\
H. Georgi, in {\em Particles and Fields -- 1974}, Proc. of the 
APS 1974 Meeting, ed. C.A. Carlson, AIP, New York, 1975. 

\bibitem{Dienes} For review, see e.g. K. Dienes, hep-th/9602045, 
and references therein. 

\bibitem{ALS} 
B. Ananthanarayan, G. Lazarides and Q. Shafi, Phys. Rev. D 44 
(1991) 1613; \\
G. Arason et al., Phys. Rev. Lett. 67 (1991) 2933. 

\bibitem{ADHRS} 
G.W. Anderson, S. Dimopoulos, L.J. Hall, S. Raby and G. Starkman, 
Phys. Rev. D 49 (1994) 3660; \\ 
K.S. Babu and Q. Shafi, Phys. Lett. B 357 (1995) 365; \\ 
K.S. Babu and R.N. Mohapatra, Phys. Rev. Lett. 74 (1995) 2418; \\  
K.S. Babu and S.M. Barr, Phys. Rev. Lett. 75 (1995) 2088; \\ 
Z. Berezhiani, Phys. Lett. B 355 (1995) 178.


\bibitem{MVM}  
S. Dimopoulous and F. Wilczek, NSF-ITP-82-07 (unpublished); \\
M. Srednicki, Nucl. Phys. B 202 (1982) 327. 

\bibitem{BB93}
K.S. Babu and S.M. Barr, Phys. Rev. D48 (1993) 5354. 

\bibitem{BB94} 
K.S. Babu and S.M. Barr, Phys. Rev. D50 (1994) 3529.

\bibitem{HR} 
L.J. Hall and S. Raby, Phys. Rev. D51 (1995) 5624. 

\bibitem{GS} 
M. Green and J. Schwarz, Phys. Lett. B 149 (1984) 117. 

\bibitem{FI} 
P. Fayet and J. Iliopoulos, Phys. Lett. 51B (1974) 461. 

\bibitem{DSW} 
M. Dine, N. Seiberg and E. Witten, Nucl. Phys. B 289 (1987) 585; \\ 
J. Atick, L. Dixon and A. Sen, {\em ibid.} B 292 (1987) 109; \\ 
M. Dine, I. Ichinose and N. Seiberg, {\em ibid.} B 293 (1987) 253. 

\bibitem{IR} 
L. Iba\~nez and G.G. Ross, Phys. Lett. B 332 (1994) 100; \\ 
P. Binetruy and P. Ramond, Phys. Lett. B 350 (1995) 49; \\ 
V. Jain and R. Shrock, {\em ibid.} B 352 (1995) 83; \\ 
E. Dudas, S. Pokorski and C. Savoy, {\em ibid.} B 356 (1995) 45; 
B 369 (1995) 255; \\ 
P. Binetruy, S. Lavignac and P. Ramond, Nucl. Phys. 
B 447 (1996) 353. 

\bibitem{Gia} 
G. Dvali and S. Pokorski, Report CERN-TH/96-286, hep-ph/9610431. 




\bibitem{IMDM} Z. Berezhiani and Z. Tavartkiladze, 
Report INFN-FE 13/96, hep-ph/9611277. 

\bibitem{Dvali} 
G. Dvali, Phys. Lett. B287 (1992) 101; 372 (1996) 113; \\
Z. Berezhiani, in ref. \cite{ADHRS}.  

\bibitem{dim5} S. Weinberg, Phys. Rev. D 26 (1982) 287; \\ 
N. Sakai and T. Yanagida, Nucl. Phys. B197 (1982) 533. 

\bibitem{Nath} 
J. Hisano, H. Murayama and T. Yanagida, Nucl. Phys. B 402 
(1993) 46; \\ 
P. Nath, A. Chamseddine and R. Arnovit, Phys. Rev. D 
32 (1985) 2348; Phys. Lett. B 287 (1992) 89. 


\bibitem{FN} 
C.D. Frogatt and H.B. Nielsen, Nucl. Phys. B 147 (1979) 277; \\ 
Z.G. Berezhiani, Phys. Lett. B 129 (1983) 99; B 150 (1985) 177; \\
S. Dimopoulos, Phys. Lett. B 129 (1983) 417; \\ 
J. Bagger, S. Dimopoulos, H. Georgi and S. Raby, in Proc. 5th 
Workshop on Grand Unification (Providence, Rhode Island 1994), 
eds. K. Kang et al., World Scientific, Singapore, 1994.  




\bibitem{PDG} Particle Data Group, Phys. Rev. D 54, Part I 
(1996) 1. 



\bibitem{seesaw} M. Gell-Mann, P. Ramond and R. Slansky, 
in {\em Supergravity}, eds. P. van Niewuenhuizen and D. Freedman 
(North-Holland, Amsterdam, 1979) p. 315; \\ 
T. Yanagida, Prog. Th. Phys. B 135 (1979) 66; \\ 
R. Mohapatra and G. Senjanovic, Phys. Rev. Lett. 44 (1980) 912. 


\bibitem{ANP} Y. Fukuda et al., Phys. Lett. B 335 (1994) 237. 

\bibitem{MSW} 
S. Mikheyev and A. Smirnov, Yad. Fiz. 42 (1995) 1441; \\ 
L. Wolfenstein, Phys. Rev. D 17 (1978) 2369. 


\end{thebibliography}
\end{document}